\documentclass[twocolumn,showpacs,preprintnumbers,amsmath,amssymb]{revtex4}

\usepackage{graphicx}
\usepackage{dcolumn}
\usepackage{bm}

\begin{document}

\preprint{APS/123-QED}

\title{ The Electrostatic Component of the Disjoining Pressure and the Pore Creation Rate in Electroporation Models and Theory}

\author{Zlatko Vasilkoski}
 \altaffiliation[Email: ]{zlatko@alumini.tufts.edu}
\affiliation{%
022 Dana Research Center, Northeastern University, Boston, MA 02115
}%


\date{\today}

\begin{abstract}
Under externally applied electric fields, lipid membranes tend to permeate and change their electrical resistance by the combined processes of pore creation and pore evolution (expansion or contraction). This study is focused on the pore creation process, represented by an empirical expression currently used in the electroporation (EP) models, for which an alternative theoretically based expression was provided. The choice of this expression was motivated by the role the DLVO's (disjoining) pressures may play in the process of EP. The electrostatic energy effects on each sides of a lipid membrane were evaluated in terms of the electrostatic component of the disjoining pressure. Thus the pore creation energy considerations in the current EP models, associated with the necessity of an idealized non conducting circular pre-pore were avoided. As a result, a new expression for the onset of the electroporation was proposed. It was found that this new theoretically determined expression is in good agreement with the one estimated from an experiment that specifically targeted the pore nucleation by eliminating the pore evolution process. Furthermore, the proposed expression is dependent on the electrolyte and the lipid properties and should provide better predictability than the currently used pore creation EP model, which is only temperature dependent.
\end{abstract}

\maketitle

\section{Introduction}

The application of electric fields to cells with intention to introduce various molecules into them is widely used physical method. This technique of permeating the cellular membrane, called electroporation (EP), has been routinely performed for many decades. By applying relatively strong electric pulses, the fluid $5$ nm thick bilipid membrane that envelops the cell and the organelles, can be perforated by nanometer size pores that are long-lived (ms) compared to the time it takes to create them (ns). It is still puzzling why the resealing period is at least six orders of magnitude longer. Most likely this is a reflection of the difference in the physics involved in these two processes. These electro pores are too small to be optically observed. The existence of these perforations in the membrane is usually confirmed by measuring the abrupt change in membrane's electrical resistance, which can be attributed to a combination of two different processes: nucleation of new pores and expansion or contraction of the existing ones (pore evolution). Experimentally, by measuring the membrane's electrical resistance, it is very hard to differentiate between these two processes. 

Although the main goal is to understand the electroporation of cells, in order to understand the EP process better, much experimental and theoretical work has been done on planar bilayer membrane. The current theoretical understanding of the electroporation process is built on the expressions originally used in the work on thin films. This is natural considering the composition similarity between thin films and planar bilayer membranes. They both consist of a two-dimensional fluid layer of lipid molecules interacting with surrounding aqueous solution. 

\section{Modeling the two aspects of the EP Process - pore nucleation and pore evolution}

The basic theoretical description of hole nucleation in thin films goes back to 1940's in the work of Kramers and Zeldovich \cite{Derjaguin_1989}, and starts with considering the expression for reversible work of formation of a pre-pore, a macroscopic hole having a circular shape. The main contributors to this mechanical energy expression are the line tension of this circular hole and the surface tension of the membrane water interface. These kinds of energy considerations are also used in EP theory and modeling to describe the two main effects an applied electric field would have on a membrane, namely creation of new pores and expansion or contraction of the existing ones.

This study does not address the process of pore evolution (expansion/contraction) in the EP models. It is focused only on the pore nucleation aspects of the EP theory and modeling.

\subsection{Currently used pore creation rate for the process of pore nucleation in the EP models}

In the current EP models the process of pore nucleation in membranes under applied electric field is described by a pore creation rate expression that besides the mechanical energy it also includes the electrostatic energy effects. Extending Litster's 1975 pore fluctuation model \cite{Litster_1975}, and Chizmadzhev 1979 work on mechanisms of defect origination in membranes \cite{AbidorEtAlChizmadzhevMainFacts1stPaper1979_3},in 1981 the following pore creation rate expression was introduced by Weaver and Mintzer \cite{WeaverMintzerEporeTheoryBilayerStabilityPhysLett1981}

\begin{eqnarray}
\frac{dN}{dt} = A_\mathrm{0} e^{- \left( \delta_\mathrm{c} - \beta\Delta\psi^2_\mathrm{m} \right)/{kT} } = \alpha e^{\beta \Delta \psi^2_\mathrm{m}/{kT}} 
\label{eq:JCW_PCR},
\end{eqnarray}
with $A_\mathrm{0} = \nu_\mathrm{0} A_\mathrm{m} d_\mathrm{m}$. Here $N$ is the number of created pores, $\nu_\mathrm{0}$ is referred to as attempt rate density per volume, $A_\mathrm{m}$ and $d_\mathrm{m}$ are membrane's area and thickness, $kT$ the Boltzmann's constant and the temperature, $\Delta\psi_\mathrm{m}$ is the transmembrane voltage, constant $\beta$ in units of capacitance is associated with the electrostatic energy effects, and $\delta_\mathrm{c}$ is the energy barrier to be overcome for a single pore to be formed. Thus $\delta_\mathrm{c}$ is the one pore formation free energy required for cooperative topological rearrangement of many lipid molecules from planar two-dimensional to most likely traversable wormhole-like three-dimensional structure. The actual principles of membrane defect formations are not known yet, but it is generally accepted that the process is cooperative and stochastic, probably involving not more than $\approx 2 \times$50 lipid molecules \cite{Vasilkoski_2006}. 

A similar version of the pore creation rate was suggested in 1988 by Glaser et al. \cite{GlaserEtAl_ElectroporationReversibleFormationEvolutionPores_BAA1988}, with $A_\mathrm{0} = {\nu A_\mathrm{m}}/{ A_\mathrm{hg}}$, where $A_\mathrm{hg}$ is the headgroup area of a lipid molecule and $\nu$ is the frequency of lateral fluctuations of the lipid molecules. This choice of parameters in $A_\mathrm{0}$ is based on the authors of \cite{GlaserEtAl_ElectroporationReversibleFormationEvolutionPores_BAA1988} believe that the rate at which the pore can open is mainly determined by lipid headgroup fluctuations. 

\subsection{Currently used pre factor $\beta$ in the EP models}

Expression (\ref{eq:JCW_PCR}) includes the electrostatic energy effects associated with the transmembrane voltage, $\Delta\psi_\mathrm{m}$, which tend to decrease the stability of the membrane against thermal fluctuations. In current EP models the way the electrostatic energy modifies the effective mechanical pore formation energy is by considering the pore as a circular capacitor. The pre factor $\beta$ (capacitance) contains the difference in energy when this capacitor is field with water instead of lipid membrane 

\begin{eqnarray}
\beta = \frac{\pi r^2_{*} \left( \varepsilon_\mathrm{w} - \varepsilon_\mathrm{m} \right)}{2d_\mathrm{m}}  
\label{eq:Beta},
\end{eqnarray}
where $r_{*}$ is some minimum pre-pore radius, and $\varepsilon_\mathrm{w}$ and $\varepsilon_\mathrm{m}$ are the corresponding dielectric constants of water and lipids. 

During the last two decades, most of the EP computer simulation models \cite{BarnettWeaverEporeUnifiedTheoryBreakdownRuptureBB1991, Weaver_Chizmadzhev_1996, DeBruinKrassowska_1998, NeuKrassowska_1999_PhysRevE, DeBruinKrassowska_1999, JoshiSchoenbach_PhysRev2000} use a general pore creation rate given by (\ref{eq:JCW_PCR}), with choice of parameters $\alpha (\nu_\mathrm{0}, A_\mathrm{m},\delta_\mathrm{c})$ and $\beta$ that provide a reasonable experimental predictability of the EP models. In some of these models \cite{NeuKrassowska_1999_PhysRevE,DeBruinKrassowska_1999}, it is taken that $\beta={1}/{\left (\Delta \psi_{ep} \right )^2}$, where $\Delta \psi_{ep}$ specifies the onset of the electroporation and is referred to as the minimal transmembrane voltage that can cause a pore to form.

With time, it is been always a theoretical challenge to accommodate the experiments demonstrating that lower and lower voltages can cause electroporation. Very often in the EP models this has been achieved by adjusting the value of the pre factor $\beta$. Since $\beta$ appears in the exponent, it strongly influences the situation that (\ref{eq:JCW_PCR}) is attempting to model. As a result, some degree of ambiguity regarding the value of $\beta$ has been present in modeling the EP process. Over the past two decades the different values of $\beta$ used in the EP models have ranged from 4.70 $kT$V$^{-2}$ to 62.50 $kT$V$^{-2}$ (at temperature $T=293$ K and $k=1.38\times 10^{-23}$ JK$^{-1}$). 

Chronologically, in 1991 \cite{BarnettWeaverEporeUnifiedTheoryBreakdownRuptureBB1991}, by assuming minimum pore radius $r_*=1$ nm and membrane thickness $d_\mathrm{m}=2.8$ nm the authors have obtained $\beta=95.70$ $kT$V$^{-2}$  but used a more realistic $\beta=4.70$ $kT$V$^{-2}$ in their EP model. By 1996 \cite{Weaver_Chizmadzhev_1996}, the authors considered more standardized values of $r_{*}=0.4$ nm and $d_\mathrm{m}=5$ nm and obtained $\beta=8.59$ $kT$V$^{-2}$. In 1998 \cite{DeBruinKrassowska_1998}, the authors have assumed $\beta=62.50$ $kT$V$^{-2}$ without providing information about the parameters used to estimate this value. The authors of \cite{NeuKrassowska_1999_PhysRevE} in 1999 have proposed that pores are being created at all radii, with decreasing probability for the larger radii, and they suggested a value $\beta=13.12$ $kT$V$^{-2}$. Similarly in their 1999 paper \cite{DeBruinKrassowska_1999}, the authors use $\beta=15.02$ $kT$V$^{-2}$. But they also assume $r_*=0.76$ nm, which would give $\beta=30.95$ $kT$V$^{-2}$. The authors of \cite{JoshiSchoenbach_PhysRev2000} also take that the pores are being created at all the radii in a decreasing manner, from their minimum pore radius $r_*=0.8$ nm, for which $\beta=34.34$ $kT$V$^{-2}$. 

An increase in the applied voltage will result in increased membrane conductance, which can be attributed to a combination of two different processes. The process of creation of new pores in the membrane and the process of expansion of the existing pores. When both of these processes are present, experimentally it is impossible to determine which part of the increased conductance is due to the newly created pores and which part is due to expansion of the existing ones. Thus relevant to the pore creation rate would be an experiment that will avoid the process of pore expansion and would only reveal the fundamentals of pore formation. 

Exactly this kind of experiment, essential for the pore creation rate was performed in 2001 \cite{Melikov_2001}, in which a nucleation of a single pore was experimentally observed. This experiment was done by applying not an electric pulse but a constant low voltage across an $A_\mathrm{m}=3$ $\mu$m $\times$ 3 $\mu$m artificial bilipid membrane patch, until a single pore appeared and resealed. Surprisingly, even voltages as low as $250$ mV could cause an appearance of a single pore. Based on this experimental data, crucial for the pore creation rate, the EP model parameters were roughly estimated in \cite{Vasilkoski_2006} to be $\alpha = 0.01$ s$^{-1}$ and $\beta=20$ $kT$V$^{-2}$ thus giving to $\alpha$ and $\beta$ in expression (\ref{eq:JCW_PCR}) some experimental justification. 

\subsection{Critique of the currently used pre factor $\beta$ in the EP models}

Further in this paper, the validity of the theoretical assumptions, based on a pre-pore existence (i.e. circular capacitor), contained in expression (\ref{eq:Beta}) are criticized and an alternative expressions for the electrostatic energy effects on pore nucleation in (\ref{eq:JCW_PCR}) is suggested.

The existing theoretical approach for the pore creation rate is based on the reversible work, assuming the existence of a minimum size pore in the membrane. This approach does not consider how the changes in the electrolyte due to the applied voltage led to the nucleation of this minimum size pore. As such, all of the so far used EP models do not take into account the electrolyte or lipid properties. Thus they give the same prediction for different types of electrolytes and counterion concentrations. It is hard to imagine that solution properties would not play any role in the EP process. This property independence hinders the predictability of the EP models and requires recurring adjustment of the model constants to fit the experiments. Since constants don't change unless they are parameters, for the predicting purposes of the EP models, it is very important to use theoretically justified parameters in their pore creation expressions. 

The current theoretical understanding of the expression (\ref{eq:Beta}) relies on the well-known fact that the lipid membrane behaves essentially as a local circular capacitor, upon the application of an electric field. Expression (\ref{eq:Beta}) is certainly valid for large holes in the membrane but geometrically it cannot be circular for small ones. In this context all current pore nucleation models assume pre existence of a circular pore with a cross section $\pi r^2_{*}$, of minimum pore radius $r_{*} \approx 0.4-1$ nm. Geometrically, this can hardly be correct, considering that the minimum radius pore is made of only few (5-6) lipid headgroups of comparable radius $\approx 0.8$ nm (area $A_\mathrm{hg}\approx 0.6$ nm$^2$ \cite{Israelachvili}), thus making $\pi r^2_{*}$ in (\ref{eq:Beta}) a very crude approximation to a circular capacitor even without including the permanent motion of the lipid molecules. In addition, the dielectric properties appearing in expression (\ref{eq:Beta}) most likely differ from the bulk phase, taken as $\varepsilon_{w}=80\varepsilon_{0}$ and $\varepsilon_{m}=2\varepsilon_{0}$ by all the EP models. 

In light of all this, a possible resolution to some of these issues related to the electrostatic energy effects in the pore creation rate is further suggested. 

\section{Application of DLVO on a lipid membrane}

For a number of decades the DLVO theory, started by Derjaguin and Landau (1941), and complemented by Verwey and Overbeek (1948), has been the well accepted theory that describes the interaction of membranes with the surrounding electrolyte, giving an expression for the (disjoining) pressure on the membrane in terms of the counterion properties. Thus it seems natural to apply DLVO to the pore creation rate used in EP models. 

\subsection{The electric potential near a lipid membrane}

\begin{figure*}
\includegraphics[scale=0.85]{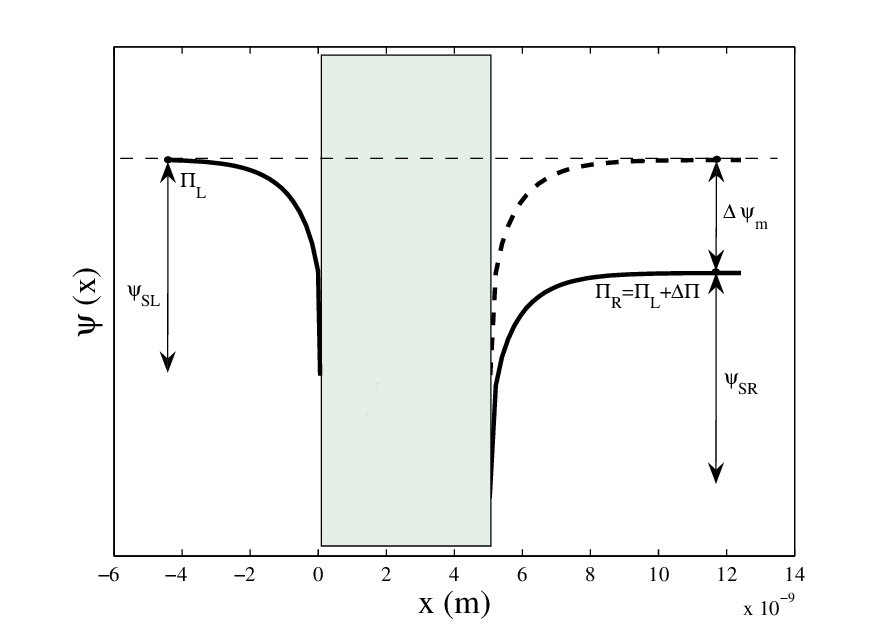}
\caption{\label{fig:epsart} Profile of the electric potentials $\it\psi\rm (x)$ near a membrane. In the absence of an external electric field the side potential is represented by the dashed line on the right side. If an external electric field is applied, it will cause an appearance of a transmembrane voltage $\Delta \psi_\mathrm{m}$ across the membrane, but it will have virtually no effect \cite{Waltz} on the side potentials, $\psi_\mathrm{SL}$ and $\psi_\mathrm{SR}$, that depend on the lipid composition (surface charge) and on the electrolyte's bulk concentration on each side of the membrane \cite{Israelachvili,Waltz}. The potential jump $\Delta \psi_\mathrm{m}$ in the bulk, according to the definition of the electrostatic component of the disjoining pressure, expression (\ref{eq:Pel}), will create a pressure $\Delta \Pi$ on the membrane that is given by expression (\ref{eq:DPel}).} \label{fig:Psi}
\end{figure*}

A membrane in electrolyte solution is screened from the environment by the electrolyte counterions that are distributed against it and they balance membrane's surface charge \cite{Israelachvili,Waltz}. The counterion concentration in the vicinity of the membrane is up to two orders of magnitude higher than in the bulk. This causes a side electric potential $\psi_\mathrm{S}$ to be formed near the membrane within few Debye lengths ($\approx 1$ nm) from it, as illustrated in Fig. \ref{fig:Psi}. The magnitude of the side potential on the left $\psi_\mathrm{SL}$ and the right side $\psi_\mathrm{SR}$ is determined by the membrane's surface charge density $\sigma$, pH, as well as the bulk ion concentration $n_0$ in the solution on each side. According to Grahame Equation \cite{Israelachvili,Waltz}, $\psi_\mathrm{S} \propto {\sigma}/{\sqrt n_0}$. For instance if the bulk concentration is increased, the side potential will get smaller. This will not result in appearance of transmembrane voltage $\Delta \psi_\mathrm{m}$ across the membrane. A typical value of the side potential is -67 mV, as given in reference \cite{Israelachvili}, estimated for a $100$ mM aqueous solution of NaCl, against a typical lipid membrane surface charge of $-0.2$ Cm$^{-2}$. There is an additional potential drop of $\approx 100$ mV in the Stern layer near the membrane \cite{Israelachvili,Derjaguin_1987, Derjaguin_1989}. 

Regardless of the membrane's surface charge or the bulk ion concentration, when no external electric field is applied, the electric potentials in the bulk on the left and on the right side are equal \cite{Israelachvili,Waltz}. This is illustrated by the dashed line in Fig. \ref{fig:Psi}. An applied electric field redistributes the charges and the potentials near the membrane and creates a transmembrane voltage ($\Delta \psi_\mathrm{m}$). Important to note \cite{Israelachvili,Waltz} is that the externally applied electric field has virtually no effect on the side potentials $\psi_\mathrm{SL}$ and $\psi_\mathrm{SR}$. This indicates that the applied electric field across a membrane manifests itself only by the transmembrane voltage \cite{Israelachvili,Waltz}.

\subsection{The electrostatic component of the disjoining pressure}

In DLVO it is customary to separate the various contributions of the disjoining pressure into different components: electrostatic, dispersion, steric, adhesive, etc. In this paper only the electrostatic $\Pi_\mathrm{e}$ component of the disjoining pressure was considered. The expression for $\Pi_\mathrm{e}$ on one side of a single membrane exposed to electrolyte of charge density $\rho$, has the following form \cite{VerweyOverbeek_1948, Derjaguin_1987,Zlatko_2006}

\begin{eqnarray}
\Pi_\mathrm{e} = -\int\limits_{0}^{\psi_\mathrm{1}} {\rho \left( \psi \right)} d\psi
\label{eq:Pel}.
\end{eqnarray}

The sign indicates that the force pushing the counterions in the electrolyte is directed towards the membrane. The integration by the potential $\psi$ is done at a particular location $x$ away from the membrane, as the potential $\psi(x)$ at that location changes from zero to some new value $\psi_\mathrm{1}$. An observation made by Derjaguin \cite{Derjaguin_1987, Derjaguin_1989} in late 1930's illustrates that the disjoining pressure $\Pi_\mathrm{e}$, after brief equilibration of the ions, is constant throughout the electrolyte (i.e. independent of the location $x$). Thus when an electric field is applied across the membrane, $\Pi_\mathrm{e}$ can conveniently be estimated by the change of $\psi$ in the bulk. 

In a symmetric situation, when same bulk concentration $n_{0}$ is present on each side of the membrane, and no external electric field is applied, the bulk electric potential is zero as illustrated by the dashed line in Fig. \ref{fig:Psi}. In this situation the membrane does not experience a difference in $\Pi_\mathrm{e}$ on each side. 

If an external electric field is applied, it breaks the symmetry in the electric potential around the membrane as illustrated by the full line in Fig. \ref{fig:Psi}. The potential in the bulk on the right increases from zero to the value of the transmembrane voltage $\Delta\psi_\mathrm{m}$. According to the definition of $\Pi_\mathrm{e}$, expression (\ref{eq:Pel}), the membrane will experience a pressure $\Delta \Pi_\mathrm{e}$ from the right side.






The actual expression for the electrostatic component of the disjoining pressure $\Pi_\mathrm{e}$, for a Boltzmann charge density distribution $\rho \left( \psi \right)$ in expression (\ref{eq:Pel}) was derived in 1948 by Verwey and Overbeek. This expression applied on one side of a single membrane exposed to 1:1 electrolyte has the following form \cite{VerweyOverbeek_1948, Derjaguin_1987,Israelachvili,Zlatko_2006}

\begin{eqnarray}
\Delta \Pi_\mathrm{e} \approx -2 n_0 kT \left. \left[\frac{1}{2} {\left(\frac{{ze}}{{kT}} \right)^2 \psi^2 } \right] \right|_{0}^{\Delta\psi_\mathrm{m}}=-n_0 \frac{{e^2 }}{{kT}} {\Delta \psi_\mathrm{m}}^2 
\label{eq:DPel},
\end{eqnarray}
where expression (\ref{eq:Pel}) was integrated over the sum of the anion and the cation charge density distribution (valence $z=\pm1$).

The asymmetry on each side of the membrane can be realized in many circumstances. For example by combination of any of these conditions: different counterions (valence), bulk concentrations $n_\mathrm{0}$, different lipid composition (surface charge) or by applying an external electric field. A detailed asymmetric analysis is beyond the scope of this work, even though some of these asymmetric conditions may be biologically relevant. In addition there is no experimental data on appearance of a single pore under asymmetric conditions. Thus further in this paper only symmetric conditions were analyzed, namely same bulk concentration and lipid composition on each side of the membrane, and a comparison was made with the symmetric experimental set up in \cite{Melikov_2001}, vital for the pore creation rate, since it led to appearance of a single pore. It also should be mentioned that the pore creation rate currently used in the EP models ignores any asymmetry since it is independent of any electrolyte properties.





\section{DLVO based expression for the pore creation rate and $\beta$ in the EP models}

Motivated by these DLVO considerations and \cite{Zlatko_2006}, in this paper, instead of the current pore creation rate given by (\ref{eq:JCW_PCR}), the following DLVO based pore creation rate is proposed  

\begin{eqnarray}
\frac{dN}{dt} = \alpha e^{{a A_\mathrm{hg} \Delta \Pi}/{kT}}
\label{eq:DLVO_PCR},
\end{eqnarray}
where $\Delta \Pi= \Pi_\mathrm{L} - \Pi_\mathrm{R}$, as Fig. \ref{fig:Psi} illustrates, is the difference in the disjoining pressure on each sides of the membrane. In the spirit of mean field theory, namely reducing the multi-body interaction to effective one-body interaction, this disjoining pressure (energy density) was multiplied by an effective volume $aA_\mathrm{hg}$ of the electrolyte, that contributed to formation of a single pore. This is the volume in which a lipid headgroup of area $A_\mathrm{hg}$ feels the disjoining pressure from the collisions with the counterions of radius $a$. 

In general, the $\Delta\Pi$ term in expression (\ref{eq:DLVO_PCR}) contains the difference of all of the disjoining pressure components. This paper was focused only on $\Delta\Pi_\mathrm{e}$ and its pre factor $\beta$. Since $\Delta\Pi$ is additive, the contribution of the other pressure components will affect only the pre factor $\alpha$. The influence of the other disjoining pressure components on $\alpha$ in (\ref{eq:DLVO_PCR}) will be analyzed in a future work. 

The general pore creation rate (\ref{eq:JCW_PCR}) was compared with the DLVO based pore creation rate (\ref{eq:DLVO_PCR}), considering only the electrostatic component of the disjoining pressure ($\Delta \Pi = -\Delta \Pi_\mathrm{e}$). Using the expression (\ref{eq:DPel}) for $\Delta \Pi_\mathrm{e}$ the value of the DVLO based pre factor $\beta$ was determined to be 

\begin{eqnarray}
\beta = n_\mathrm{0} \frac {e^2}{kT} a A_\mathrm{hg}
\label{eq:B_PCR}.
\end{eqnarray}

Unlike the previously used expression (\ref{eq:Beta}) for $\beta$ which requires existence of a circular pre-pore and is independent of the electrolyte properties, the pore creation rate with the new pre factor (\ref{eq:B_PCR}) is dependent on the electrolyte and lipid properties such as counterion size $a$, bulk concentration $n_\mathrm{0}$ and lipid headgroup size $A_\mathrm{hg}$. In this context, the value of $\beta=20$ $kT$V$^{-2}$ that was estimated in \cite{Vasilkoski_2006} is specific to the experimental setup of \cite{Melikov_2001}, namely $100$ mM bulk concentration of K$^\mathrm{+}$ counterions. An estimate of the DLVO based expression (\ref{eq:B_PCR}), at room temperature $T=293$ K and with counterion radius of $a_\mathrm{K^+}=3.3$ $\mathrm{\AA}$ \cite{Israelachvili}, gave a value of $\beta=18.67$  $kT$V$^{-2}$. This value, based entirely on theoretical grounds, is very close to the round, experimentally estimated one in \cite{Vasilkoski_2006}, bearing in mind that in \cite{Vasilkoski_2006}, $\beta$ was estimated in conjunction with $\alpha$ as a $\chi^2$ fit to the experimental data.

\section{Conclusion}

In conclusion, based on DLVO theoretical consideration, an alternative pore creation rate expression was proposed for use in the EP models. The energy effects in the proposed expression are due to the difference in the disjoining pressure $\Delta \Pi$ on each sides of the membrane. More specifically, the electrostatic energy effects on a lipid membrane were considered in terms of the electrostatic component of the disjoining pressure $\Delta \Pi_\mathrm{e}$, and a new expression for $\beta$ was suggested that avoids the necessity of an idealized circular pre-pore. The value of this new $\beta$, founded entirely on theoretical arguments, was established to be in a good agreement with the estimated experimental value in \cite{Vasilkoski_2006}. Since it is dependent on the electrolyte and the lipid properties, this DLVO based $\beta$ should provide better predictability than the current EP models that regard $\beta$ as a constant. The role that other components of the disjoining pressure may play in the EP theory and how they may affect $\alpha$ in (\ref{eq:JCW_PCR}) will be considered in a future work.

\bibliography{PCT}

\begin{thebibliography}{18}
\expandafter\ifx\csname natexlab\endcsname\relax\def\natexlab#1{#1}\fi
\expandafter\ifx\csname bibnamefont\endcsname\relax
  \def\bibnamefont#1{#1}\fi
\expandafter\ifx\csname bibfnamefont\endcsname\relax
  \def\bibfnamefont#1{#1}\fi
\expandafter\ifx\csname citenamefont\endcsname\relax
  \def\citenamefont#1{#1}\fi
\expandafter\ifx\csname url\endcsname\relax
  \def\url#1{\texttt{#1}}\fi
\expandafter\ifx\csname urlprefix\endcsname\relax\def\urlprefix{URL }\fi
\providecommand{\bibinfo}[2]{#2}
\providecommand{\eprint}[2][]{\url{#2}}

\bibitem[{\citenamefont{Derjaguin}(1989)}]{Derjaguin_1989}
\bibinfo{author}{\bibfnamefont{B.~V.} \bibnamefont{Derjaguin}},
  \emph{\bibinfo{title}{Theory of stability of colloids and thin films}}
  (\bibinfo{year}{1989}).

\bibitem[{\citenamefont{Litster}(1975)}]{Litster_1975}
\bibinfo{author}{\bibfnamefont{J.}~\bibnamefont{Litster}},
  \bibinfo{journal}{Phys.\ Lett.} \textbf{\bibinfo{volume}{53A}},
  \bibinfo{pages}{193} (\bibinfo{year}{1975}).

\bibitem[{\citenamefont{Chizmadzhev et~al.}(1979)\citenamefont{Chizmadzhev,
  Arakelyan, and Pastushenko}}]{AbidorEtAlChizmadzhevMainFacts1stPaper1979_3}
\bibinfo{author}{\bibfnamefont{Y.~A.} \bibnamefont{Chizmadzhev}},
  \bibinfo{author}{\bibfnamefont{V.~B.} \bibnamefont{Arakelyan}},
  \bibnamefont{and} \bibinfo{author}{\bibfnamefont{V.~F.}
  \bibnamefont{Pastushenko}}, \bibinfo{journal}{Bioelectrochem. Bioenerget.}
  \textbf{\bibinfo{volume}{6}}, \bibinfo{pages}{63} (\bibinfo{year}{1979}).

\bibitem[{\citenamefont{Weaver and
  Mintzer}(1981)}]{WeaverMintzerEporeTheoryBilayerStabilityPhysLett1981}
\bibinfo{author}{\bibfnamefont{J.~C.} \bibnamefont{Weaver}} \bibnamefont{and}
  \bibinfo{author}{\bibfnamefont{R.~A.} \bibnamefont{Mintzer}},
  \bibinfo{journal}{Phys. Lett.} \textbf{\bibinfo{volume}{86A}},
  \bibinfo{pages}{57} (\bibinfo{year}{1981}).

\bibitem[{\citenamefont{Vasilkoski et~al.}(2006)\citenamefont{Vasilkoski,
  Esser, Gowrishankar, and Weaver}}]{Vasilkoski_2006}
\bibinfo{author}{\bibfnamefont{Z.}~\bibnamefont{Vasilkoski}},
  \bibinfo{author}{\bibfnamefont{A.~T.} \bibnamefont{Esser}},
  \bibinfo{author}{\bibfnamefont{T.~R.} \bibnamefont{Gowrishankar}},
  \bibnamefont{and} \bibinfo{author}{\bibfnamefont{J.~C.}
  \bibnamefont{Weaver}}, \bibinfo{journal}{Phys. Rev. E}
  \textbf{\bibinfo{volume}{74}}, \bibinfo{pages}{021904}
  (\bibinfo{year}{2006}).

\bibitem[{\citenamefont{Glaser et~al.}(1988)\citenamefont{Glaser, Leikin,
  Chernomordik, Pastushenko, and
  Sokirko}}]{GlaserEtAl_ElectroporationReversibleFormationEvolutionPores_BAA19%
88}
\bibinfo{author}{\bibfnamefont{R.~W.} \bibnamefont{Glaser}},
  \bibinfo{author}{\bibfnamefont{S.~L.} \bibnamefont{Leikin}},
  \bibinfo{author}{\bibfnamefont{L.~V.} \bibnamefont{Chernomordik}},
  \bibinfo{author}{\bibfnamefont{V.~F.} \bibnamefont{Pastushenko}},
  \bibnamefont{and} \bibinfo{author}{\bibfnamefont{A.~I.}
  \bibnamefont{Sokirko}}, \bibinfo{journal}{Biochim. Biophys. Acta}
  \textbf{\bibinfo{volume}{940}}, \bibinfo{pages}{275} (\bibinfo{year}{1988}).

\bibitem[{\citenamefont{Barnett and
  Weaver}(1991)}]{BarnettWeaverEporeUnifiedTheoryBreakdownRuptureBB1991}
\bibinfo{author}{\bibfnamefont{A.}~\bibnamefont{Barnett}} \bibnamefont{and}
  \bibinfo{author}{\bibfnamefont{J.~C.} \bibnamefont{Weaver}},
  \bibinfo{journal}{Bioelectrochem. and Bioenerg.}
  \textbf{\bibinfo{volume}{25}}, \bibinfo{pages}{163} (\bibinfo{year}{1991}).

\bibitem[{\citenamefont{Weaver and
  Chizmadzhev}(1996)}]{Weaver_Chizmadzhev_1996}
\bibinfo{author}{\bibfnamefont{J.~C.} \bibnamefont{Weaver}} \bibnamefont{and}
  \bibinfo{author}{\bibfnamefont{A.}~\bibnamefont{Chizmadzhev}},
  \bibinfo{journal}{Bioelectrochem. Bioenerget.} \textbf{\bibinfo{volume}{41}},
  \bibinfo{pages}{135} (\bibinfo{year}{1996}).

\bibitem[{\citenamefont{De{B}ruin and
  Krassowska}(1998)}]{DeBruinKrassowska_1998}
\bibinfo{author}{\bibfnamefont{K.~A.} \bibnamefont{De{B}ruin}}
  \bibnamefont{and}
  \bibinfo{author}{\bibfnamefont{W.}~\bibnamefont{Krassowska}},
  \bibinfo{journal}{Ann. Biomed. Eng.} \textbf{\bibinfo{volume}{26}},
  \bibinfo{pages}{584} (\bibinfo{year}{1998}).

\bibitem[{\citenamefont{Neu and
  Krassowska}(1999)}]{NeuKrassowska_1999_PhysRevE}
\bibinfo{author}{\bibfnamefont{J.~C.} \bibnamefont{Neu}} \bibnamefont{and}
  \bibinfo{author}{\bibfnamefont{W.}~\bibnamefont{Krassowska}},
  \bibinfo{journal}{Phys. Rev. E} \textbf{\bibinfo{volume}{59}},
  \bibinfo{pages}{3471} (\bibinfo{year}{1999}).

\bibitem[{\citenamefont{De{B}ruin and
  Krassowska}(1999)}]{DeBruinKrassowska_1999}
\bibinfo{author}{\bibfnamefont{K.~A.} \bibnamefont{De{B}ruin}}
  \bibnamefont{and}
  \bibinfo{author}{\bibfnamefont{W.}~\bibnamefont{Krassowska}},
  \bibinfo{journal}{Biophys. J.} \textbf{\bibinfo{volume}{77}},
  \bibinfo{pages}{1225} (\bibinfo{year}{1999}).

\bibitem[{\citenamefont{Joshi and
  Schoenbach}(2000)}]{JoshiSchoenbach_PhysRev2000}
\bibinfo{author}{\bibfnamefont{R.~P.} \bibnamefont{Joshi}} \bibnamefont{and}
  \bibinfo{author}{\bibfnamefont{K.~H.} \bibnamefont{Schoenbach}},
  \bibinfo{journal}{Phys. Rev. E} \textbf{\bibinfo{volume}{62}},
  \bibinfo{pages}{1025} (\bibinfo{year}{2000}).

\bibitem[{\citenamefont{Melikov et~al.}(2001)\citenamefont{Melikov, Frolov,
  Shcherbakov, Samsonov, and Chizmadzhev}}]{Melikov_2001}
\bibinfo{author}{\bibfnamefont{K.~C.} \bibnamefont{Melikov}},
  \bibinfo{author}{\bibfnamefont{V.~A.} \bibnamefont{Frolov}},
  \bibinfo{author}{\bibfnamefont{A.}~\bibnamefont{Shcherbakov}},
  \bibinfo{author}{\bibfnamefont{A.~V.} \bibnamefont{Samsonov}},
  \bibnamefont{and} \bibinfo{author}{\bibfnamefont{Y.~A.}
  \bibnamefont{Chizmadzhev}}, \bibinfo{journal}{Biophys. J.}
  \textbf{\bibinfo{volume}{80}}, \bibinfo{pages}{1829} (\bibinfo{year}{2001}).

\bibitem[{\citenamefont{Israelachvili}(1991)}]{Israelachvili}
\bibinfo{author}{\bibfnamefont{J.}~\bibnamefont{Israelachvili}},
  \emph{\bibinfo{title}{Intermolecular and Surface Forces}}
  (\bibinfo{year}{1991}).

\bibitem[{\citenamefont{Waltz et~al.}(2004)\citenamefont{Waltz, Teissi\'{e},
  and Milazzo}}]{Waltz}
\bibinfo{editor}{\bibfnamefont{D.}~\bibnamefont{Waltz}},
  \bibinfo{editor}{\bibfnamefont{J.}~\bibnamefont{Teissi\'{e}}},
  \bibnamefont{and} \bibinfo{editor}{\bibfnamefont{G.}~\bibnamefont{Milazzo}},
  eds., \emph{\bibinfo{title}{Bioelectrochemistry of Membranes}}
  (\bibinfo{publisher}{Birkh\"{o}user Verlag}, \bibinfo{address}{PO Box 133,
  CH-4010 Basel, Switzerland}, \bibinfo{year}{2004}).

\bibitem[{\citenamefont{Derjaguin et~al.}(1987)\citenamefont{Derjaguin,
  Churaev, and Muller}}]{Derjaguin_1987}
\bibinfo{author}{\bibfnamefont{B.~V.} \bibnamefont{Derjaguin}},
  \bibinfo{author}{\bibfnamefont{N.~V.} \bibnamefont{Churaev}},
  \bibnamefont{and} \bibinfo{author}{\bibfnamefont{V.~M.}
  \bibnamefont{Muller}}, \emph{\bibinfo{title}{Surface Forces}}
  (\bibinfo{publisher}{Consultants Bureau}, \bibinfo{address}{233 Spring St.
  New York, NY 10013}, \bibinfo{year}{1987}).

\bibitem[{\citenamefont{Verwey and Overbeek}(1948)}]{VerweyOverbeek_1948}
\bibinfo{author}{\bibfnamefont{E.~J.~W.} \bibnamefont{Verwey}}
  \bibnamefont{and} \bibinfo{author}{\bibfnamefont{J.~T.~G.}
  \bibnamefont{Overbeek}}, \bibinfo{journal}{Elsevier}  (\bibinfo{year}{1948}).

\bibitem[{\citenamefont{Vasilkoski}(2006)}]{Zlatko_2006}
\bibinfo{author}{\bibfnamefont{Z.}~\bibnamefont{Vasilkoski}}
  (\bibinfo{year}{2006}), \eprint{arXiv:physics/0701013v1}.

\end{thebibliography}

\end{document}